\documentclass[twocolumn, prl]{revtex4-1}
\makeatletter
\newcommand*{\rom}[1]{\expandafter\@slowromancap\romannumeral #1@}
\makeatother
\usepackage{graphicx}
\usepackage{amsmath}
\usepackage{amssymb}
\usepackage{dcolumn}
\usepackage{float}
\usepackage{bm}
\usepackage{tikz}
\usepackage{booktabs} 
\usepackage{siunitx} 
\usepackage{pgfplotstable} 
\pgfplotsset{every axis/.append style={mark options={scale=0.7},
       legend style={ font=\small, mark options={scale=0.7} }, }}
\usepgfplotslibrary{units}
\usepackage[breaklinks=true,colorlinks,citecolor=blue,linkcolor=blue,urlcolor=blue]{hyperref}

\DeclareMathAlphabet{\bi}{OML}{cmm}{b}{it}
\def\be{\begin{equation}}
\def\ee{\end{equation}}
\def\bearr{\begin{eqnarray}}
\def\eearr{\end{eqnarray}}
\def\la{\langle}
\def\ra{\rangle}

\begin{document}
\title{Study of adiabatic connection in density functional theory with an accurate wavefunction for 2-electron atoms}
\bigskip
\author{Rabeet Singh Chauhan }
\email[E-mail: ]{rabeet@iitk.ac.in}
\author{Manoj K. Harbola}
\email[E-mail: ]{mkh@iitk.ac.in}
\normalsize
\affiliation
{Department of Physics, Indian Institute of Technology-Kanpur,
Kanpur-208 016, India}
\date{\today}
\begin{abstract}
Using an accurate semi-analytic wavefunction for two electron atoms, we construct the external potential $v_{ext}^{\alpha}(\vec{r})$ for varying strength of $\alpha V_{ee} (0 \leq \alpha \leq 1 )$ where $\alpha$ is the strength parameter and $V_{ee}$ is electron-electron interaction. Using this potential we explicitly calculate the energy of their positive ion and show that the ionization-potential of these systems remains unchanged with respect to $\alpha$.
Furthermore, using total energies $E^{\alpha}$ of these systems as a function of $\alpha$, we provide new perspective into a variety of hybrid functionals. 
\end{abstract}
\maketitle
\section{Introduction}
Adiabatic connection\cite{Jones} has played an important role in basic understanding of ground-state density functional theory\cite{Hohn}. In adiabatic connection (AC), a fully interacting many-electron system described by the Hamiltonian (atomic units are used throughout the letter)
\begin{equation}
H = \sum_i \Big(\frac{1}{2}\nabla_i^{2} + v_{ext}(\vec{r_i})\Big)+ \frac{1}{2} \sum_{i \neq j} \frac{1}{|\vec{r_i}-\vec{r_j}|} \label{eq1}
\end{equation}
is mapped to the corresponding Kohn-Sham\cite{Kohn} system given by the Hamiltonian 
\begin{equation}
  H_{KS} = \sum_i \Big(-\frac{1}{2}\nabla_i^2 + v_{ext}(\vec{r_i}) + v_{H}(\vec{r_i}) + v_{xc} (\vec{r_i})\Big), \label{eq2}
\end{equation}
where $v_{H}(\vec{r_i})$ and $v_{xc} (\vec{r_i})$ are the Hartree and exchange-correlation potentials. This is done by scaling  the electron-electron interaction by a parameter $\alpha$ as $V_{ee}^{\alpha} = \frac{1}{2}\sum_{i\neq j}\frac{\alpha}{|\vec{r_i}-\vec{r_j}|} (0\leq \alpha \leq 1)$ and changing $\alpha$ from $1$ (fully interacting system) to $\alpha=0$ (Kohn-Sham system) while keeping the density equal to the ground state density. To keep the density fixed, the external potential $v_{ext}(\vec{r_i})$ of Eq. (\ref{eq1}) is changed to $v_{ext}^{\alpha}(\vec{r_i})$. Thus the corresponding wavefunction $\Psi^{\alpha} (\vec{r_1}, \vec{r_2},..\vec{r_N})$ also changes but the density 
\begin{equation}
  \rho(\vec{r})= N \int |\Psi^{\alpha} (\vec{r_1}=\vec{r}, \vec{r_2},..\vec{r_N}|^2 \mathrm{d} \vec{r_2} \mathrm{d} \vec{r_3}..\mathrm{d} \vec{r_N}           \label{eq3}
\end{equation} 
remains equal to the true ground state density.  It is understood that $\Psi^{\alpha=1} (\vec{r_1}, \vec{r_2},..\vec{r_N})$ is the true many electron wavefunction while $\Psi^{\alpha=0} (\vec{r_1}, \vec{r_2},..\vec{r_N})$ is the Slater determinant formed from the Kohn-Sham orbitals.  
\par
The exchange-correlation energy in density functional theory is defined through the AC as\cite{Jones, DCP1, Gun1, Gun2, DCP2, Becke}
\begin{equation}
E_{xc}^{DFT} = \int_{0}^{1} E_{xc}^{\alpha} \mathrm{d}\alpha \label{eq4}
\end{equation}
where
\begin{equation}
E_{xc}^{\alpha}=\langle \Psi^{\alpha}|V_{ee}|\Psi^{\alpha}\rangle - \frac{1}{2}\int \frac{\rho(\vec{r})\rho(\vec{r'})}{|\vec{r}-\vec{r'}|} \mathrm{d} \vec{r} \mathrm{d} \vec{r'} \nonumber
\end{equation}
It is well known that the difference in $E_{xc}^{DFT}$ and $E_{xc}^{\alpha=1}$ is equal to the difference $T_c$ in the true kinetic energy and the Kohn-Sham kinetic energy $T_s$. Eq. (\ref{eq4}) has been used extensively to obtain hybrid exchange-correlation functionals\cite{Becke, Beck, Perd} by mixing $\langle \Psi^{\alpha=0}|V_{ee}|\Psi^{\alpha=0}\rangle$ and an approximation for exchange-correlation energies, usually the LDA\cite{Yang} or GGA\cite{GGA}.
\par
 Although AC is often cited in density functional theory studies, not much work has been done to explicitly construct $v_{xc}^{\alpha}(\vec{r})$ and $E_{xc}^{\alpha}$ and study its fundamental aspects. The initial work in this direction is that done by J. Katriel et. al.\cite{Katr} where the density is kept constant by constraining the moments of local one-body operators. Further work has been done by Teal et. al.\cite{Teal2} by using Legendre transformation on the energy functional. In this method, the functional\cite{Lieb} $$F[\rho]=\max_{v_{ext}^{\alpha}}\big[E[v_{ext}^{\alpha}]-\int \rho(\vec{r}) v_{ext}^{\alpha}(\vec{r}) d\vec{r}\big]$$ is used to obtain $v_{ext}^{\alpha}(\vec{r})$. For a given density $\rho(\vec{r})$ this is done by maximizing $$E[v_{ext}^{\alpha}]-\int \rho(\vec{r}) v_{ext}^{\alpha}(\vec{r}) d\vec{r}$$  by varying $v_{ext}^{\alpha}(\vec{r})$.  To do this $v_{ext}^{\alpha}(\vec{r})$ is expressed as a sum of Gaussians. This method has again been used to study the adiabatic connection for atoms and molecules having up to $10$ electrons at different lavels of approximations\cite{teal3}. Similar work\cite{teal4} has also been done on range-separated functionals.  In contrast, the present work uses direct approach based on Levy's constrained search method\cite{CS} and employs a simple but accurate semi-analytic wavefunction for two electron system. 
\par  
   In the work described here, we use an accurate variational form of interacting wavefunction\cite{Sech} combined with constrained-search approach to obtain $v_{ext}^{\alpha}(\vec{r})$ and $\Psi^{\alpha}(\vec{r_1}, \vec{r_2})$ for two electron  atoms for the full range of $\alpha~~ (0\leq \alpha \leq 1)$. These are then employed to show by explicit calculations that the chemical potential for a given density remains the same and is equal to the negative of the ionization potential of the true system irrespective of the value of $\alpha$. To the best of our knowledge this is the first calculation of this kind using the constrained search approach of Levy and  shows the accuracy of our study. The difference between the present work and that refs. \cite{Teal2,teal3,teal4} is that in our work search is made over the wavefunction space keeping the density constant. In the works of refs. \cite{Teal2,teal3,teal4} the search is over the space of different one-body potentials.  Having obtained $\Psi^{\alpha}(\vec{r_1}, \vec{r_2})$ we employ it to calculate the exchange-correlation energy $E_{xc}^{\alpha}[\rho]$ for the full range of $\alpha$. This is then used to explicitly calculate $E_{xc}^{DFT}$ using Eq. (\ref{eq4}) and show that the difference indeed comes out to be $T_c$ to a high degree of accuracy. More importantly,  by plotting $E_{xc}^{\alpha}$ against $\alpha$  and comparing the resulting graph to that for hybrid functionals, new perspective is provided to understand the latter.
\section{The Wavefunction}
  The ground state wavefunction that  we employ is based on the Le Sech wavefunction\cite{Sech} for 2-electron atoms and has the form
\begin{equation}
\Psi(\vec{r_1}, \vec{r_2}) = \phi(r_1) \phi(r_2) f(r_{12})
\end{equation}
where $r_{12}=|\vec{r_1}-\vec{r_2}|$ and $f(r_{12})=[\cosh(a r_1)+\cosh(a r_2)][1+0.5 r_{12} e^{-b r_{12}}]$ for the ground state. In this wavefunction $a$ and $b$ are the variational parameters and for each set of $(a, b)$, $\phi(r)$ is obtained by solving a self-consistent equation given in ref. \cite{Rabi,Baber}. The most accurate wavefunction is given for the set $(a, b)$ that minimizes the total energy.  The resulting energies, densities and the exchange-correlation potential obtained from the wavefunction are all very close to their exact values. More significantly, the wavefunction $\Psi^{\alpha}(\vec{r_1}, \vec{r_2})$ can be easily adapted to represent a many-electron wavefunction for $\alpha \neq 1$. This is given as 
\begin{equation}
\Psi^{\alpha}(\vec{r_1}, \vec{r_2}) = \phi(r_1) \phi(r_2) f^{\alpha}(r_{12})
\end{equation} 
where $$f^{\alpha}(r_{12})=[\cosh(a r_1)+\cosh(a r_2)][1+0.5 \alpha r_{12} e^{-b r_{12}}].$$ To study adiabatic connection, $\Psi^{\alpha}(\vec{r_1}, \vec{r_2})$ should be such that it gives the same density as the true interacting system density 
$\rho^{\alpha=1}(\vec{r})$. In the present work this is enforced by the Zhao-Parr\cite{Zhao} method by demanding that the integral  
\begin{equation}
  \frac{1}{2} \int \int \frac{[\rho^{\alpha}(\vec{r})-\rho^{\alpha=1}(\vec{r})][\rho^{\alpha}(\vec{r'})-\rho^{\alpha=1}(\vec{r'})]}{|\vec{r}-\vec{r}'|} \mathrm{d} \vec{r} \mathrm{d} \vec{r'}
\end{equation}
where
\begin{equation}
\rho^{\alpha}(\vec{r})=2\int |\Psi^{\alpha}(\vec{r}, \vec{r_2})|^{2} \mathrm{d} \vec{r}_2
\end{equation}
vanish.  Minimizing the expectation value of 
$$H'=\sum_i \Big(\frac{1}{2}\nabla_i^{2} + v_{ext}(\vec{r}_i)\Big)+ \frac{1}{2} \sum_{i \neq j} \frac{\alpha}{|\vec{r}_i-\vec{r}_j|}$$ with $v_{ext}(\vec{r}_i)=-\frac{Z}{r_i}$ and enforcing the constraint above with  Lagrange multiplier $\lambda$ leads to the following equation for $\Psi^{\alpha}(\vec{r}_1, \vec{r}_2)$
\begin{widetext}
\begin{eqnarray}  \nonumber
 -\frac{1}{2}\nabla^{2}\phi(r)&-&\frac{1}{A(r)}\nabla\phi(r)\cdot\int\mathrm{f^{\alpha}(r, r_{2}, r_{02})|\phi(r_{2})|^{2}\nabla f^{\alpha}(r, r_{2}, r_{02})}\mathrm{d}\vec{r}_2\\  \nonumber
&-&\frac{1}{2 A(r)}\int[\mathrm{|\phi(r_{2})|^{2}f^{\alpha}(r, r_{2}, r_{02})\nabla^{2} f^{\alpha}(r, r_{2}, r_{02})}
+\mathrm{|\phi(r_{2})|^{2}f^{\alpha}(r, r_{2}, r_{02})\nabla_{2}^{2} f^{\alpha}(r, r_{2}, r_{02})} \\ \nonumber
~~~~~~~&+&\mathrm {|f^{\alpha}(r, r_{2}, r_{02})|^{2}\phi(r_{2})\nabla_{2}^{2}\phi(r_{2})}+\mathrm{2 f^{\alpha}(r, r_{2}, r_{02})\phi(r_{2})\nabla_{2}\phi(r_{2})\cdot\nabla_{2}f^{\alpha}(r, r_{2}, r_{02})}]\mathrm{d}\vec{r}_2\phi(r) \\    \nonumber
&-&\frac{Z}{r}\phi(r)-\frac{Z}{A(r)}\int\mathrm{\frac{|\phi(r_{2})f^{\alpha}(r, r_{2}, r_{02})|^{2}}{r_{2}}}\mathrm{d}\vec{r}_2 \phi(r)      
+\frac{\textcolor{blue}{\alpha}}{A(r)}\int\frac{|\phi(r_{2})f^{\alpha}(r, r_{2}, r_{02})|^{2}}{|\vec{r}-\vec{r}_2|}\mathrm{d}\vec{r}_2\phi(r)\\
&+&\textcolor{blue}{\lambda}[v_{zp}(r) 
 +\frac{1}{A(r)}\int {v_{zp}(r_2)|\phi(r_2)f^{\alpha}(r_2,r,r_{20})|^2}\mathrm{d}\vec{r}_2
]\phi(r)=E_2^{\alpha} \phi(r) \label{maineq} 
\end{eqnarray}
\end{widetext}
here $$ r_{0i}=|\vec{r}-\vec{r}_i|,~A(r) = \int |\phi(r_2)f^{\alpha}(r,r_2,r_{02})|^2 \mathrm{d}\vec{r}_2$$
and  
\begin{equation}
v_{zp}(r)=\int \frac{[\rho^{\alpha}(r_2)-\rho^{\alpha=1}(r_2)]}{|\vec{r}-\vec{r}_{2}|}\mathrm{d}\vec{r}_2.
\end{equation}
where $\rho^{\alpha}(\vec{r}_2)$ is given by Eq.(\ref{eq3}) using $\phi(r)$ obtained from Eq. (\ref{maineq}) to construct $\Psi^{\alpha}(\vec{r_1}, \vec{r_2})$. As in the case of fully interacting system, for each set of $(a, b)$ the equation above is solved self-consistently. The eigenvalue $E_2^{\alpha}$ of Eq. (\ref{maineq}) is the energy of the two-electron system for a given ($a, b$). The appropriate $\Psi^{\alpha}(\vec{r_1}, \vec{r_2})$ is given by that $(a, b)$ that leads to minimum value of $E_2^{\alpha}$ for a large value of $\lambda$. In our calculations we have chosen $\lambda=1000$. Going beyond $\lambda=1000$ does not change the resulting values by any significant amount. The resulting $\lambda v_{zp}(r)$ is the difference between $v_{ext}^{\alpha=1}=-\frac{Z}{r}$ and $v_{ext}^{\alpha}(\vec{r})$.
\section{Computational aspects}
 To facilitate the calculations we make use of the following property of $v_{ext}^{\alpha}(\vec{r})$
 \begin{equation}
 v_{ext}^{\alpha}(\vec{r})\rightarrow \frac{-Z+(1-\alpha)(N-1)}{r}~ as~ r\rightarrow \infty.
 \end{equation}
This has been proved\cite{Colon} in the past on the basis of the behavior of the exchange potential that goes as $-\frac{\alpha}{r}$ as $r \rightarrow \infty$. We give an alternative argument here using the asymptotic decay of the density which is related to the ionization-potential\cite{Osten,Katr2,LPS,PPLB} $I$ or the chemical potential ($\mu=-I$) of the system as $$\rho(r\rightarrow \infty) \sim e^{-2\sqrt{2 I}r}.$$
Since during the adiabatic connection density is kept unchanged, this leads to the energy difference between the energy of a system and its ion is equal to $-I$ irrespective of the value of $\alpha$. This then indicates that\cite{Barth}
\begin{equation}
\lim_{r\rightarrow \infty} v_{ext}^{\alpha}(\vec{r}) + \lim_{r\rightarrow \infty} \sum_{j=1}^{N-1} \frac{\alpha}{|\vec{r}-\vec{r_j}|} \rightarrow \frac{-Z+N-1}{r}. 
\end{equation}
Since $$\lim_{r\rightarrow \infty} \sum_{j=1}^{N-1} \frac{\alpha}{|\vec{r}-\vec{r_j}|} \rightarrow \frac{\alpha(N-1)}{r},$$ this implies that 
\begin{equation}
  \lim_{r\rightarrow \infty} v_{ext}^{\alpha}(\vec{r}) + \frac{\alpha(N-1)}{r} \rightarrow \frac{-Z+N-1}{r} 
\end{equation}
or $ v_{ext}^{\alpha}(\vec{r}) \rightarrow \frac{-Z+(1-\alpha)(N-1)}{r}$ in this limit. On the other hand, near the nucleus the potential $v_{ext}^{\alpha}(\vec{r})$ goes as $-\frac{Z}{r}$ because the resulting density should satisfy the cusp condition\cite{Kato}.  Thus in carrying out the numerical calculations in Eq. (\ref{maineq}) we use the potential 
\begin{equation}
  -\frac{Z}{r} + (1-\alpha)(1-\frac{1}{N}) v_{H}(\vec{r}) \nonumber
  \end{equation}
in place of $-\frac{Z}{r}$, where $v_{H}(\vec{r}) = \int \frac{\rho^{\alpha=1}(\vec{r}')}{|\vec{r}-\vec{r}'|}(\vec{r})$. Note that the term with $v_{H}(\vec{r})$  has been included for modified ee-interaction along with self-interaction term subtracted from it. Furthermore, for $\alpha \rightarrow 0$ and $\alpha \rightarrow 1$ this term goes to the appropriate forms. The external potential $v_{ext}^{\alpha}(\vec{r})$ is then given as 
\begin{eqnarray}
v_{ext}^{\alpha}(\vec{r}) = -\frac{Z}{r} &+& (1-\alpha)(1-\frac{1}{N}) v_{H}(\vec{r})\\
                                         & +& \lambda \int \frac{[\rho^{\alpha}(\vec{r}')-\rho^{\alpha=1}(\vec{r}')]}{|\vec{r}-\vec{r}'|}\mathrm{d}\vec{r}'.   \nonumber
\end{eqnarray} \\
Although in the past the Zhao-Parr method has been used extensively\cite{Morr,mkhs1,mkhs2} for constructing Kohn-Sham system, this is the first time that it is being applied to Hamiltonian containing e-e interaction term. 
\section{Results} 
Now we present the results of our calculations for He atom. The results for other two-electron atoms are similar in nature.\\\\ 
(\rom{1})\textbf{Energy and chemical potentials as function of $\alpha$} : 
 We first give the results for the energy and chemical potential for the He atom. For different $\alpha$, the corresponding values of $a$ and $b$ parameter are given in Table \ref{tb1} along with the energies $E_2^{\alpha}$. This energy as a function of $\alpha$ is plotted in Fig. (\ref{fig1}) where  $E_{2}^{\alpha=0}=-1.8078$, which is the sum of the Kohn-Sham eigenvalues and $E_2^{\alpha=1}=-2.9028$. More importantly, energies for all other values of $\alpha$ are also plotted and it is seen that $E_2^{\alpha}$ as a function of $\alpha$ is essentially linear. We have also calculated the energy $E_{1}^{\alpha}$ of a single electron in potential $v_{ext}^{\alpha}(\vec{r})$. These are also given in Table \ref{tb1} and plotted against $\alpha$ in Fig. (\ref{fig1}). These energies also vary linearly, being equal to the Kohn-Sham eigenvalue for $\alpha=0$ and $-\frac{Z^2}{2}$ for $\alpha=1$. As shown in Table (\ref{tb1}), the difference $E_2^{\alpha}-E_1^{\alpha}$ is a constant and is equal to negative of the experimental ionization potential\cite{Lide} (0.9037[a.u.]) of He to an excellent degree of accuracy. This is also depicted in Fig. (\ref{fig1}). This is the first explicit calculation of ionization potential for different $\alpha's$ and demonstration of its constancy. The results also show the precision of our calculations.
\squeezetable
\begin{table}[H]
\caption{Energies  $E_2^{\alpha}$ and $E_1^{\alpha}$ for He atom calculated for different $\alpha$ at constant density. Here $a$ and $b$ are paramters in the correlated part of the wavefunction at which we get the minimum energy $E_2^{\alpha}$. Also given is the chemical potential $\mu=E_2^{\alpha}-E_1^{\alpha}$. \label{tb1}}
\begin{ruledtabular}
\begin{tabular}{cccccc}
$\alpha$   &   a  &   b  & $E_2^{\alpha}$ & $E_1^{\alpha}$ & $\mu$  \\ \hline
 0.0    & 0.00 & 0.00 &  -1.8078 & -0.9039 &  -0.9039\\ 
 0.1    & 0.36 & 0.18 &  -1.9112 & -1.0074 &  -0.9039\\
 0.2    & 0.46 & 0.18 &  -2.0164 & -1.1126 &  -0.9038\\
 0.3    & 0.52 & 0.20 &  -2.1231 & -1.2194 &  -0.9037\\
 0.4    & 0.57 & 0.18 &  -2.2314 & -1.3278 &  -0.9036\\
 0.5    & 0.64 & 0.18 &  -2.3408 & -1.4374 &  -0.9034\\
 0.6    & 0.69 & 0.18 &  -2.4515 & -1.5482 &  -0.9033\\
 0.7    & 0.74 & 0.18 &  -2.5633 & -1.6600 &  -0.9033\\
 0.8    & 0.79 & 0.18 &  -2.6769 & -1.7734 &  -0.9035\\
 0.9    & 0.85 & 0.20 &  -2.7894 & -1.8862 &  -0.9032\\
 1.0    & 0.93 & 0.20 &  -2.9031 & -2.0000 &  -0.9030     
\end{tabular}
\end{ruledtabular}
\end{table}  
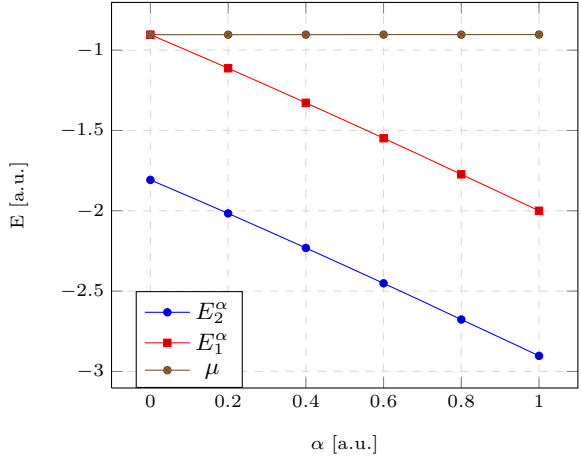
\begin{figure}
  \begin{center}
    \caption{ Total energy $E_2^{\alpha}$, single electron energy $E_1^{\alpha}$ for He and the corresponding energy difference $E_2^{\alpha}-E_1^{\alpha}$ for different values of $\alpha$. We note that the energy curves are almost linear and parallel to each other so the chemical potential $\mu$ is constant for all the values of $\alpha$.}
 \label{fig1}
  \vspace{0.5cm}
    \begin{tikzpicture}
      \begin{axis}[
          width=0.9\linewidth, 
          grid=major, 
          grid style={dashed,gray!30}, 
          xlabel=$\alpha$, 
          ylabel=E,
          x unit={a.u.}, 
          y unit={a.u.},
          mark repeat = {2},
          legend style={at={(axis cs:0.1,-2.5)},anchor=north}, 
          x tick label style={rotate=0,anchor=north} 
        ]
        \addplot 
        table[x=column 0,y=column 1, col sep=comma] {he_e_vs_al.csv};
        \addplot
        table[x=column 0,y=column 2, col sep=comma] {he_e_vs_al.csv};
        \addplot
        table[x=column 0,y=column 3, col sep=comma] {he_e_vs_al.csv};
        \legend{$E_2^{\alpha}$,$E_1^{\alpha}$,$\mu$}
      \end{axis}
    \end{tikzpicture}
  \end{center}
\end{figure}
(\rom{2}) \textbf{Correction to the potential $-\frac{Z}{r}$}: Plotted in Fig. (\ref{fig2}) is the potential term $v_{\lambda}^{\alpha}(\vec{r})=v_{ext}^{\alpha}(\vec{r})+\frac{Z}{r}$, which is given as 
\begin{equation}
 v_{\lambda}^{\alpha}(\vec{r}) = \lambda \int \frac{\rho^{\alpha}(\vec{r'})-\rho^{\alpha=1}(\vec{r'})}{|\vec{r}-\vec{r'}|} \mathrm{d} \vec{r'}\\ +  (1-\alpha)(1-\frac{1}{N}) v_{H}(\vec{r}) \label{eq16} 
\end{equation}
This correction is plotted for $\alpha=0$ (Kohn-Sham system), $0.3$, $0.6$, $0.9$ and $\alpha=1$ (the true system). For $\alpha=0$, the term in Eq. (\ref{eq16}) is the Hartree plus the exchange-correlation potential for the He atom. It is seen that the structure of correction to $-\frac{Z}{r}$
remains similar for all values of $\alpha$ but its magnitude changes and becomes equal to zero for $\alpha=1$. It is also interesting to look all the term $\lambda \int \frac{\rho^{\alpha}(\vec{r'})-\rho^{\alpha=1}(\vec{r'})}{|\vec{r}-\vec{r'}|} \mathrm{d} \vec{r'}$ which is given in Fig. (\ref{fig3}). From the magnitude of this term which is two orders of magnitude smaller than that of $v_{\lambda}^{\alpha}(\vec{r})$, it is clear that the main difference between $-\frac{Z}{r}$ and $v_{ext}^{\alpha}(\vec{r})$ arises from the Hartree potential scaled appropriately to take care of the e-e interaction and self interaction of an electron.

\begin{figure}
  \begin{center}
    \caption{Potential $v_{\lambda}^{\alpha}(r)$ (Eq. \ref{eq16})needed to be added to the potential $-\frac{Z}{r}$to keep the density constant while changing the electron-electron interaction. The curves are plotted for He ($Z=2$).}
    \label{fig2}
    \begin{tikzpicture}
      \begin{semilogxaxis}[
          scale only axis,
          width=0.7\linewidth, 
          grid=major, 
          grid style={dashed,gray!30}, 
          xlabel=$r$, 
          ylabel=$v_{\lambda}^{\alpha}(r)$,
          x unit={a.u.}, 
          y unit={a.u.},
          mark repeat={50},
          x tick label style={rotate=90,anchor=east} 
        ]
        \addplot
        table[x=column 0,y=column 1, col sep=comma] {v_lam_al.csv};
        \addplot
        table[x=column 0,y=column 2, col sep=comma] {v_lam_al.csv};
        \addplot
        table[x=column 0,y=column 3, col sep=comma] {v_lam_al.csv};
        \addplot
        table[x=column 0,y=column 4, col sep=comma] {v_lam_al.csv};
        \addplot
        table[x=column 0,y=column 5, col sep=comma] {v_lam_al.csv};
        \legend{$\alpha=0.0$,$\alpha=0.3$,$\alpha=0.6$,$\alpha=0.9$,$\alpha=1.0$}
      \end{semilogxaxis}
    \end{tikzpicture}
  \end{center}
\end{figure}
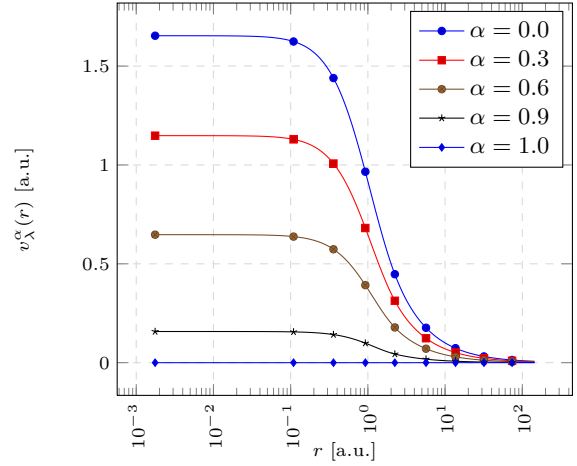
\begin{figure}
  \begin{center}
    \caption{The first term $\lambda \int \frac{\rho^{\alpha}(\vec{r'})-\rho^{\alpha=1}(\vec{r'})}{|\vec{r}-\vec{r'}|} \mathrm{d} \vec{r'}$ of $v_{\lambda}^{\alpha}(r)$ (Eq. \ref{eq16}) is plotted here. The curves are plotted for He ($Z=2$).}
    \label{fig3}
    \begin{tikzpicture}
      \begin{semilogxaxis}[
          scale only axis,
          width=0.8\linewidth, 
          grid=major, 
          grid style={dashed,gray!30}, 
          xlabel=$r$, 
          ylabel=$v_{\lambda}^{\alpha}(r)$,
          x unit={a.u.}, 
          y unit={a.u.},
          mark repeat={50},
          x tick label style={rotate=90,anchor=east} 
        ]
        \addplot
        table[x=column 0,y=column 1, col sep=comma] {v_lam_al_F.csv};
        \addplot
        table[x=column 0,y=column 2, col sep=comma] {v_lam_al_F.csv};
        \addplot
        table[x=column 0,y=column 3, col sep=comma] {v_lam_al_F.csv};
        \addplot
        table[x=column 0,y=column 4, col sep=comma] {v_lam_al_F.csv};
        \addplot
        table[x=column 0,y=column 5, col sep=comma] {v_lam_al_F.csv};
        \legend{$\alpha=0.0$,$\alpha=0.3$,$\alpha=0.6$,$\alpha=0.9$,$\alpha=1.0$}
      \end{semilogxaxis}
    \end{tikzpicture}
  \end{center}
\end{figure}
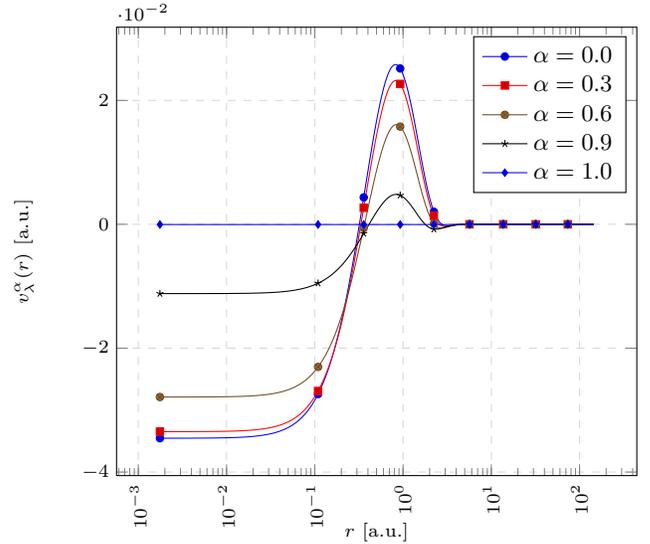
(\rom{3})\textbf{$E_{xc}^{DFT}$ from Hellmann-Feynman  theorem} :
 Exchange-correlation energy in density functional theory is given by Eq. (\ref{eq4}). Displayed in Table \ref{tb2} are values of $E_{xc}^{\alpha}$ for different values of $\alpha$. It is seen that with increasing $\alpha$, the value of $E_{xc}^{\alpha}$ becomes more negative. This has been plotted in Fig. (\ref{fig4}) where graph of $E_{xc}^{\alpha}$ versus $\alpha$ is almost a straight line. The behavior shown in the figure is similar to that given in ref.\cite{teal3}. Numerical integration $\int E_{xc}^{\alpha} \mathrm{d} \alpha$ gives a value of $E_{xc}^{DFT}=-1.0682 au$, while $E_{xc}^{\alpha=1}=-1.1036au$. This gives $T_c=E_{xc}^{DFT}-E_{xc}^{\alpha=1}=0.0354 au$ which is very close to the actual $T_c=T^{\alpha=1}-T^{KS(\alpha=0)}=0.0370au$. This again shows the correctness of wavefunction employed by us.
 \begingroup
\squeezetable
\begin{table}[h]
\caption{Expectation values $\la V_{ee} \ra^{\alpha}$ calculated for different wavefunctions at different $\alpha$ at constant density. By subtracting the Hartree energy $E_H$ from $\la V_{ee} \ra$.\label{tb2}}
\begin{ruledtabular}
\begin{tabular}{cccc}
$\alpha$&  $\la V_{ee} \ra^{\alpha}$ & $E_{H}$ & $E_{xc}^{\alpha}=\la V_{ee} \ra^{\alpha}-E_{H}$  \\ \hline
 0.0   &  1.0245  &  2.0490 & -1.0245  \\ 
 0.1   &  1.0138  &  2.0490 & -1.0351  \\
 0.2   &  1.0039  &  2.0490 & -1.0450  \\
 0.3   &  0.9954  &  2.0490 & -1.0535  \\
 0.4   &  0.9873  &  2.0490 & -1.0617  \\
 0.5   &  0.9796  &  2.0490 & -1.0694  \\
 0.6   &  0.9722  &  2.0490 & -1.0768  \\
 0.7   &  0.9652  &  2.0490 & -1.0837  \\
 0.8   &  0.9565  &  2.0490 & -1.0925  \\
 0.9   &  0.9500  &  2.0490 & -1.0990  \\
 1.0   &  0.9454  &  2.0490 & -1.1036       
\end{tabular}
\end{ruledtabular}
\end{table} \\ \\

 (\rom{4})\textbf{Study of Hybrid Functionals} :
 Starting with Becke, various hybrid functionals have been proposed over the past $25$ years.  The idea behind these functional is to mix an appropriate ratio of the exchange-correlation energy $E_{xc}^{\alpha=0}$ calculated in terms of Kohn-Sham orbitals ( the expression is the same as that of Hartree-Fock exchange energy in terms of orbitals) and the approximate energy $E_{xc}^{\alpha=1}$ for $\alpha=1$.  Thus the exchange-correlation energy is given as $$E_{xc}^{Hybrid}=\gamma E_{xc}^{\alpha=0} + (1-\gamma)E_{xc}^{\alpha=1},$$ where $\gamma$ is an appropriately chosen fraction. This can be thought of in terms of adiabatic-connection formula as a linear function of $\alpha$
 \begin{equation}
   E_{xc}^{\alpha} = (1-n \alpha) E_{xc}^{\alpha=0} + n\alpha E_{xc}^{\alpha=1}. \label{eq17}
 \end{equation} 
 We do this interpolation keeping in mind the way $E_{xc}^{\alpha}$ behaves with $\alpha$, it is very close to being linear.
 Note that for $\alpha\rightarrow 0$ and $\alpha\rightarrow 1$ the interpolation above has the correct limits. Performing the integral $\int_{0}^{1} E_{xc}^{\alpha} \mathrm{d} \alpha$ and approximating $E_{xc}^{\alpha=1}$ with density functional approximation (DFA) leads to
\begin{equation}
(1-\frac{n}{2}) E_{xc}^{\alpha=0}+\frac{n}{2} E_{xc}^{\alpha=1}(DFA).
\end{equation} 
Thus $n=2(1-\gamma)$. We now compare three hybrid-functional  viz. $E_{xc}^{Becke}$, $E_{xc}^{PBE}$ and $E_{xc}^{B3LYP}$ for the exact density $\rho^{\alpha=1}(\vec{r})$. These functionals are given as\cite{Beck, Perd, Beck1, LYP}
\begin{eqnarray*}
E_{xc}^{Becke}&=& \frac{1}{2} E_{xc}^{\alpha=0} + \frac{1}{2} E_{xc}^{LDA} \\
E_{xc}^{PBE0} &=& \frac{1}{4} E_{xc}^{\alpha=0} + \frac{3}{4} E_{x}^{PBE} + E_{c}^{PBE}\\
E_{xc}^{B3LYP}&=& E_{xc}^{LDA}+0.2(E_{xc}^{\alpha=0}-E_x^{LDA})\\
                         &~&+0.72(E_x^{GGA}-E_x^{LDA}) \\
                         &~&+0.82(E_c^{GGA}-E_c^{LDA})
\end{eqnarray*}

 We parametrize them as linear functions of $\alpha$ as follows
\begin{eqnarray}
  E_{xc}^{Becke}(\alpha) &=& (1-\alpha) E_{xc}^{\alpha=0} + \alpha E_{xc}^{LDA}  \nonumber \\
  E_{xc}^{PBE0}(\alpha)   &=& (1-\frac{3\alpha}{2}) E_{xc}^{\alpha=0} + \frac{3\alpha}{2} E_{x}^{PBE} + E_{c}^{PBE} \nonumber \\
  E_{xc}^{B3LYP}(\alpha) &=&(1-1.6\alpha)E_{xc}^{\alpha=0}+2 \alpha E_{xc}^{LDA}-0.4\alpha E_x^{LDA} \nonumber\\
                         &~&+1.44\alpha(E_x^{GGA}-E_x^{LDA}) \nonumber \\
                         &~&+1.62\alpha(E_c^{GGA}-E_c^{LDA}) \label{eq19}
\end{eqnarray} 
\begin{figure}
  \begin{center}
    \caption{Exchange-correlation energy $E_{xc}^{Hybrid}(\alpha)$ versus $\alpha$ for functionals given in Eq. (\ref{eq19}) and (\ref{20}).}
    \label{fig4}
    \begin{tikzpicture}
      \begin{axis}[
          width=\linewidth, 
          grid=major, 
          grid style={dashed,gray!30}, 
          xlabel=$\alpha$, 
          ylabel=$E_{xc}^{Hybrid}(\alpha)$,
          x unit={a.u.}, 
          y unit={a.u.},
          mark repeat = {2},
          legend style={at={(axis cs:0.9,-1.02)},anchor=north},
          x tick label style={rotate=0,anchor=north} 
        ]
        \addplot
        table[x=column 0,y=column 1, col sep=comma] {he_hyb.csv};
        \addplot
        table[x=column 0,y=column 5, col sep=comma] {he_hyb.csv};
        \addplot
        table[x=column 0,y=column 2, col sep=comma] {he_hyb.csv};
        \addplot
        table[x=column 0,y=column 3, col sep=comma] {he_hyb.csv};
        \addplot
        table[x=column 0,y=column 4, col sep=comma] {he_hyb.csv};
        \legend{Present,B3LYP,Becke,PBE0,PBE96}
      \end{axis}
    \end{tikzpicture}
  \end{center}
\end{figure}
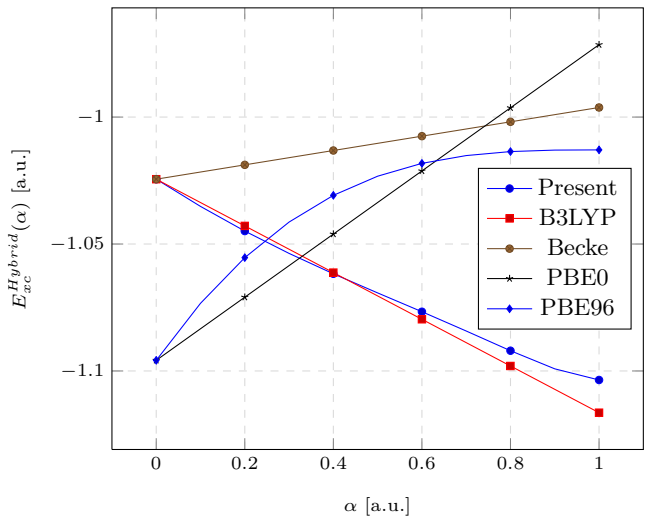
These are plotted as a function of $\alpha$ in Fig. (\ref{fig4}) and compared with the exact $E_{xc}^{\alpha}$. In addition we also plot 
\begin{equation}
E_{xc}^{PBE96}(\alpha) = (E_{xc}^{\alpha=0} - E_{x}^{PBE})(1-\alpha)^3 + (E_{c}^{PBE}+E_{x}^{PBE}) \label{20}
\end{equation}
 as was proposed by Perdew et al. \cite{Perd}. Notice that both $E_{xc}^{PBE0}(\alpha)$ and $E_{xc}^{PBE96}(\alpha)$ lead to the same functional upon integration over $\alpha$. However, $PBE0$ and $PBE96$ differ at $\alpha=1$. The corresponding exchange-correlation energies given by these approximations are, respectively, $E_{xc}^{Becke}=-1.0135$, $E_{xc}^{PBE0}=-1.0337$, $E_{xc}^{B3LYP}=-1.0705$ while the exact DFT exchange-correlation energy $E_{xc}^{DFT}=-1.0682$.  Fig. (\ref{fig4}) provides insights into why Becke and PBE0 functionals underestimate the magnitude of the true exchange-correlation energy while B3LYP is quite accurate. While the functional $E_{xc}^{Becke}(\alpha)$ is always smaller in magnitude that the exact $E_{xc}^{\alpha}$, the $PBE0$ and $PBE96$ functionals cross the $E_{xc}^{\alpha0}$ curve and that leads to cancellation of errors. On the other hand, $B3LYP$ functional follows the $E_{xc}^{\alpha}$ curve closely.  However, we note that for extended systems $B3LYP$ does not give accurate results because of its failure to reproduce homogeneous electron gas limit\cite{Perd1}. We note that in the past $B3LYP$ functional has been compared with the exact $E_{xc}^{\alpha}$ and our work confirms the previous results\cite{teal3}. However we have also shown how other functionals behave in comparision to the exact behavior. We finally mention that in the past an attempt has been made to model $E_{xc}^{\alpha}$ by Peach et. al.\cite{Teal1}. 
\section{Concluding Remarks}
To conclude, we have presented a comprehensive study of adiabatic connection using an accurate two-electron wavefunction in conjunction with the constrained search method. The accuracy of our study is indicated by the fact that the chemical potential remains constant as a function of the adiabatic connection parameter $\alpha$ and the kinetic energy component $T_{c}$ is also determined precisely through Hellmann-Feynman theorem applied with the wavefunction used. We have constructed the external potential for different values of $\alpha$  and shown that the major component of it comes from the Hartree potential for a given ground-state density. Furthermore, we have studied different hybrid functionals as a function of $\alpha$ and shed light on their behavior as a function of $\alpha$. This may help in designing better hybrid-functionals. We note that recently a hybrid exchange-correlation functional has also been proposed by mixing $E_{xc}^{\alpha=0}$ and $E_{xc}^{\alpha=\infty}$ limits\cite{Ern}. A study of this functional along the lines given here will be taken up in the future.


\begin{thebibliography}{60}
\bibitem{Jones}
J. Harris and R. O. Jones, J.  Phys F $\mathbf{4}, 1170(1974)$
\bibitem{Hohn}
P. Hohenberg and W. Kohn, Phys. Rev. $\mathbf{136}, B864(1964)$
\bibitem{Kohn}
W. Kohn and L. J. Sham, Phys. Rev. $\mathbf{140}, A1133(1965)$
\bibitem{DCP1}
D. C. Langreth and J. P. Perdew, Solid State Commun. $\mathbf{17}, 1425(1975)$
\bibitem{Gun1}
O. Gunnarsson and B. I. Lundqvist, Phy. Rev. B $\mathbf{13}, 4274(1976)$
\bibitem{Gun2}
O. Gunnarsson and B. I. Lundqvist, Phy. Rev. B $\mathbf{15}, 6006(1977)$
\bibitem{DCP2}
D. C. Langreth and J. P. Perdew, Phy. Rev. B $\mathbf{15}, 2884(1977)$
\bibitem{Becke}
A. D. Becke, J.  Chem.  Phys. $\mathbf{140}, 18A301(2014)$ 
\bibitem{Beck}
A.  D.  Becke, J.  Chem.  Phys. $\mathbf{98}, 1372(1993)$
\bibitem{Perd}
J. P. Perdew, M. Ernzerhof and K. Burke, J.  Chem.  Phys. $\mathbf{105}, 9982(1996)$
\bibitem{Yang}
R. G. Parr and W. Yang, \emph{Density Functional Theory of Atoms and Molecules} (Oxford, New York, 1989)
\bibitem{GGA}
J. P. Perdew, K. Burke, M. Ernzerhof, Phys.  Rev.  Lett. $\mathbf{77}, 3865(1996)$
\bibitem{Katr}
J. Katriel, S. Roy and M. Springborg, J.  Chem.  Phys. $\mathbf{121}, 12179(2004)$ 
\bibitem{Teal2}
A. M. Teale, S. Coriani and T. Helgaker, J. Chem. Phys. $\mathbf{130}, 104111(2009)$
\bibitem{Lieb} 
E.  H.  Lieb, Int. J. Quantum Chem. $\mathbf{24},243(1983)$ 
\bibitem{teal3}
A. M. Teale, S. Coriani and T. Helgaker, J. Chem. Phys. $\mathbf{132}, 164115(2010)$
\bibitem{teal4}
A. M. Teale, S. Coriani and T. Helgaker, J. Chem. Phys. $\mathbf{133}, 164112(2010)$   
\bibitem{CS}
M. Levy, Proc. Natl. Acad. Sci. USA $\mathbf{76}, 6062(1979)$
\bibitem{Sech}
C Le Sech, J. Phys. B: Atom. Mol. Opt. Phys. $\mathbf{30},L47(1997)$
\bibitem{Rabi}
R.  S.  Chauhan and M. K. Harbola, Chem.  Phys.  Lett.  $\mathbf{639}, 248(2015)$
\bibitem{Baber}
T.  D.  H.  Baber and H. R. Hasse, Math. Proc. of Cambridge Philosophical Soc.  $\mathbf{33}, 253(1937)$
\bibitem{Zhao}
Q.  Zhao and R.  G.  Parr, J.  Chem.  Phys.  $\mathbf{98}, 543(1993)$
\bibitem{Colon}
F. Colonna and A. Savin, J.  Chem.  Phys. $\mathbf{110}, 2828(1999)$
\bibitem{Osten}
M. Hoffman-Ostenhof and T. Hoffman-Ostenhof, Phys. Rev. A $\mathbf{16}, 1782(1977)$
\bibitem{Katr2}
J. Katriel and E. R. Davidson, Proc. Natl. Acad. Sci. $\mathbf{77}, 4403(1980)$
\bibitem{PPLB}
J. P. Perdew, R. G. Parr, M. Levy and J. L. Balduz, Jr. ,  Phys. Rev. Lett. $\mathbf{49}, 1691(1982)$
\bibitem{LPS}
M. Levy, J. P. Perdew and V. Sahni, Phys. Rev. A $\mathbf{30}, 2745(1984)$
\bibitem{Barth}
C.  -O.  Almbladh and U.  von Barth, Phys.  Rev.  B  $\mathbf{31}, 3231(1985)$
\bibitem{Kato}
T. Kato, Commun. Pure Appl. Math. $\mathbf{10}, 151(1957)$
\bibitem{Morr}
R. C. Morrison and Q. Zhao, Phys. Rev. A $\mathbf{51}, 1980(1995)$
\bibitem{mkhs1}
P. Samal, Manoj K. Harbola and A. Holas, Chem. Phys. Lett. $419, 217(2006)$
\bibitem{mkhs2}
P. Samal and Manoj K. Harbola, J. Phys. B: At. Mol. Opt. Phys. $39, 4065(2006)$
\bibitem{Lide}
D.R. Lide, CRC Handbook of Chemistry and Physics, 2006, pp. 1526
\bibitem{Lee}
R. van Leeuwen, O. Gritsenko and E. J. Baerends, Z. Phys. D $\mathbf{33}, 229(1995)$
\bibitem{Beck1}
A.  D.  Becke, Phys. Rev. A $\mathbf{38}, 3098(1988)$
\bibitem{Beck2}
A.  D.  Becke, J.  Chem.  Phys. $\mathbf{98}, 5624(1993)$
\bibitem{LYP}
C. Lee, W. Yang and R. G. Parr, Phys.  Rev.  B $\mathbf{37}, 785(1988)$
\bibitem{Perd1}
J. P. Perdew and K. Burke, Int. J. Quantum Chem. $\mathbf{57}, 309(1996)$
\bibitem{Teal1}
Michael J. G. Peach, A. M. Miller, A. M. Teale and D. J. Tozer, J. Chem. Phys. $\mathbf{129}, 064105(2008)$
\bibitem{Ern}
Y. Zhou, H. Bahmann and M. Ernzerhof, J. Chem. Phys. $\mathbf{143}, 124103(2015)$
%
\end{thebibliography}
\end{document}